\documentclass[suppldata]{interact}
\usepackage{multirow}
\pdfoutput=1

\usepackage[doublespacing]{setspace}

\usepackage[longnamesfirst,sort]{natbib}
\bibpunct[, ]{(}{)}{;}{a}{,}{,}
\renewcommand\bibfont{\fontsize{11}{12}\selectfont}
\usepackage{makecell}

\theoremstyle{plain}

\theoremstyle{definition}

\theoremstyle{remark}

\begin{document}

\title{Evaluating the Cauchy Combination Test for Count Data}
\author{
\name{Huda Alsulami\textsuperscript{a,b}\thanks{CONTACT Huda Alsulami. Email: haalsulami@kau.edu.sa} 
Silvia Liverani\textsuperscript{a}}
\affil{\textsuperscript{a}Centre for Probability, Statistics and Data Science, School of Mathematical Sciences, Queen Mary University of London, London, England, E1 4NS, United Kingdom; \textsuperscript{b} Department of Statistics, King Abdulaziz University, Jeddah, Makkah, 21589, Saudi Arabia.}
}

\maketitle

\begin{abstract}
The Cauchy combination test (CCT) is a $p$-value combination method used in multiple-hypothesis testing and is robust under dependence structures. This study aims to evaluate the CCT for independent and correlated count data where the individual $p$-values are derived from tests based on Normal approximation to the negative binomial distribution. The correlated count data are modelled via copula methods. The CCT performance is evaluated in a simulation study to assess the type 1 error rate and the statistical power and compare it with existing methods. In addition, we consider the influence of factors such as the success parameter of the negative binomial distribution, the number of individual $p$-values, and the correlation strength. Our results indicate that the negative binomial success parameter and the number of combined individual $p$-values may influence the type 1 error rate for the CCT under independence or weak dependence. The choice of copula method for modelling the correlation between count data has a significant influence on type 1 error rates for both the CCT and MinP tests. The CCT has more control over managing the type 1 error rate as the strength increases in the Gumbel-Hougaard copula. This knowledge may have significant implications for practical applications.   
\end{abstract}

\begin{abbreviations}
CCT, Cauchy combination test; MinP, minimum $P$-value test
\end{abbreviations}

\begin{keywords}
Cauchy combination test; Copula methods; count data; multiple testing; negative binomial distribution
\end{keywords}

\section{Introduction}
Combining $p$-values from various statistical tests is a fundamental procedure in multiple testing for applied statistics. It is a tool to detect an overall effect, such as in meta-analysis and bioinformatics. These combination tests combine and unify large numbers of $p$-values to a single $p$-value, potentially providing a more powerful test than testing each $p$-value separately. Suppose there are $m$ hypotheses to be tested simultaneously. Let $H_{0i}$ and $H_{ai}$ represent the null and alternative hypotheses for the $ith$ variable, respectively, where $i = 1,\ldots, m$. Let $\textbf{\textit{T}} = (T_1, \ldots, T_m)^T$ be the vector of test statistics corresponding to the $m$ hypotheses, along with their associated $p$-values $\textbf{\textit{P}} = (p_1, \ldots, p_m)^T$.
The purpose of the $p$-values combination test is testing: 
\begin{equation*}
      H_0:\bigcap_{i=1}^{m}H_{0i}
      \hspace{8mm}
      \text{versus} 
     \hspace{8mm}
     H_a: \bigcup_{i=1}^{m}H_{ai}
\end{equation*}
where $H_0$, the global null hypothesis, is satisfied when all the individual null hypotheses $H_{0i}$ are true and $H_a$, the global alternative hypothesis, if at least one of the individual alternatives $H_{ai}$ is true. Combining multiple tests provides a comprehensive conclusion about a specific research question. Moreover, it improves statistical power and controls the inflation of type 1 errors. These $p$-value combination methods inherently account for the number of combined tests, thereby avoiding the need for multiple testing corrections.
\par
In the literature, many $p$-value combination tests, which differ by their underlying assumptions, have been proposed to combine independent individual $p$-values \citep{Fisher1932, Stouffer1949}. Extensions of these combination tests have been developed to include dependence and weights \citep{Liptak1958, Lancaster1961, Brown1975, KOST2002183, Pooleetal2016}. The significance of the global alternative hypothesis may be significantly affected when the test statistic of the combined $p$-values does not appropriately account for the correlations among the individual $p$-values. It is crucial to investigate the impact of the correlation on the significance of the combined test \citep{alves2014}. For an overview of these methods see \citep{Loughin2004, alves2014, Heard2018, ZhangWu2022}. Recently, there has been interest in the Cauchy combination test (CCT) \citep{LiuXie2020} due to its advantageous features over other methods in addressing challenges arising from correlations, computations, and sparse signals in high-dimensional settings. 
\par
This study set out to evaluate the CCT’s performance for count data. The objective is to investigate the type 1 error rate and the statistical power of the CCT when the individual $p$-values are derived from test statistics based on the Normal approximation to the negative binomial distribution. The study offers some important insights by studying the influence of the negative binomial parameter, the success parameter, and the number of individual $p$-values on the combination test. Moreover, it discusses the implications on the power in the case of independent and correlated data. 
\par
This paper is organized as follows. Section~\ref{Methods} reviews three $p$-value combination methods: Fisher, MinP, and the Cauchy combination tests, and introduces the copula methods to construct correlations between count variables. The simulation study is described in Section~\ref{Simulation}, then results and discussion are presented in Section~\ref{Results}. Finally, a conclusion is given in Section~\ref{Conclusion}.  

\section{Methods}\label{Methods}
In this section, we briefly review the CCT and two other $p$-value combination tests, the MinP and Fisher's tests. In addition, we introduce the copula methods for producing correlated data.
\subsection{\textbf{$P$}-value combination methods}
\subsubsection{MinP test}
The minimum $p$-value test (MinP) \citep{Tippett1931}, orders the individual $p$-values in ascending order $p_{(1)}<p_{(2)}<\ldots<p_{(m)}$. Under the assumption that they are independent and identically uniform on the unit interval [0,1], the MinP test is $p_{(1)}$ which follows a beta distribution with parameters 1 and $m$, and its $p$-value is calculated as $1-(1-p_{(1)})^m$.
\subsubsection{Fisher's combination test}
The Fisher's combination test \citep{Fisher1932} combines the non-linear transformations of the $m$ $p$-values where each transformation $-2\log(p_i)$, under the null hypothesis, has a chi-square distribution with $2$ degrees of freedom. Therefore, the Fisher's test statistic has the following distribution:
\begin{equation*}
    \Psi_F=\sum_{i=1}^{m}-2\log(p_i) \sim \chi^2_{2m}.
\end{equation*} 
\subsubsection{Cauchy combination test}
The Cauchy combination test (CCT) by \citet{LiuXie2020} possesses advantageous features over other tests. The CCT is robust against correlation structure and powerful against sparse signals. In addition, it is computationally efficient which makes it suitable for high dimensional data analysis. The CCT is the sum of weighted non-linear $p$-values transformations through the tangent function. Under the null hypothesis, the CCT is defined as:
\begin{equation*}
\Psi_{CCT}=\sum_{i=1}^{m}\omega_i \tan\{(0.5-p_i)\pi\},
\end{equation*}
where $\omega_i>0$ and $\sum_{i=1}^{m}\omega_i=1$. If no prior information is provided, the CCT becomes the weighted average of the transformations where $\omega_i=\frac{1}{m}$. However, if the weights are random variables and independent of the individual test statistics, then the tail approximation still holds. Under various correlation structures, the correlation has a minimal impact on the tails of the CCT distribution. The tails of the CCT distribution are approximately standard Cauchy, and its $p$-value is:
 \begin{equation*}
 P_{CCT} = 0.5 - \frac{\arctan(\psi_{CCT})}{\pi}. 
 \label{PvalueCCT}
\end{equation*}
It has been theoretically and empirically demonstrated in \citep{LiuXie2020} that the CCT can effectively control type 1 error rates across different significance levels. The ratio of the size of the CCT, which represents the type 1 error rate, to the significance level approaches $1$ as the significance level converges to $0$. This indicates its validity in large-scale multiple testing.
\par
Under the global null hypothesis, the tail probability of the CCT is approximated by a standard Cauchy distribution, which is valid under the assumptions of bivariate Normality of the individual tests and some regularity conditions on the correlation matrix. \citet{Longetal2023} broaden the applicability of the CCT when the assumption of bivariate Normality may not hold. They demonstrated that the approximation of the standard Cauchy distribution for the tail probability of the CCT is still valid across a broader range of bivariate distributions. This includes the six popular copula distributions which are commonly used to model dependencies between random variables.
\subsection{Copula methods}
We utilise copulas to simulate correlated count data. Copula methods are powerful tools to capture complex dependencies between variables rather than simple linear relationships. They model various structures of dependencies, including tails dependencies. A copula is a multivariate distribution function that models the dependence structure between multiple variables, each following a standard uniform marginal distribution, U$(0,1)$ \citep{Hofert}. The basic theorem of copula theory is known as Sklar's Theorem \citep{Sklar1959}. Two types of Archimedean parametric copulas for asymmetric dependencies are considered: the Clayton and the Gumbel-Hougaard copulas, which are widely used in many applications \citep{Hofert}. The Clayton copula models positive dependence in the lower tail, while the Gumbel-Hougaard copula models positive dependence in the upper tail. They are defined by:
\begin{equation}
C(\boldsymbol{u}; \theta) = \left(1-m+ \sum_{i=1}^{m} u_i^{-\theta}\right)^{-\frac{1}{\theta}}, \quad \theta > 0, \quad \boldsymbol{u} \in [0, 1]^m 
\label{eq:Clayton}
\end{equation}
and
\begin{equation}
C(\boldsymbol{u}; \theta) = \exp\left\{-\left[\sum_{i=1}^{m} (-\log u_i)^{\theta}\right]^{1/\theta}\right\}, \quad \theta \in [1, \infty), \quad \boldsymbol{u} \in [0, 1]^m.
\label{eq:Gumbel}
\end{equation}
Both copulas have the parameter $\theta$, representing the tail dependence coefficient. As $\theta$ increases, the strength of dependence increases. 
\subsection{Evaluating the Cauchy combination test for count data}
Biological data, such as RNA-sequencing data, are best fitted with the negative binomial distribution. Unlike the Poisson distribution, the negative binomial distribution accounts for overdispersion when the variance exceeds the mean. Therefore, the CCT is adopted as a gene-set test to identify differentially expressed genes \citep{LIU2019410}. Another application of the CCT on count data, in a comparative study evaluating different methods for analysing microbiome data \citep{HamPark2022}, the CCT outperforms other methods of combining $p$-values and provides an accurate $p$-value while controlling type one error rate. In addition, the ranked combined $p$-values produced from the CCT have high-rank similarity with the true ranks. The CCT successfully replicated and identified microbiome taxa associated with colorectal cancer in a real dataset where the most highly ranked microbiome taxa using the CCT have been reported to be associated with this condition. Consequently, evaluating the CCT is pivotal to providing a robust statistical tool for analyzing non-Gaussian count data.

We aim to evaluate the type 1 error rate and the power of the CCT to combine individual $p$-values obtained from the Normal approximation to the negative binomial distribution and modelling the correlation between the discrete data via copula methods. There are several formulations for the negative binomial distribution in the literature. In this paper, we used the following definition. In a sequence of independent Bernoulli trials, the negative binomial distribution is the distribution of the number of trials (or failures) $X$ needed until a fixed number ($r$) of successes occurs. Then, $X\sim NB(r,p)$ where the first parameter $r$ is the number of successes, and $p$ is the probability of success in each trial, and the probability mass function of $X$ is:
\begin{equation}
f(x)= \binom{x+r-1}{r-1}p^r(1-p)^x, \hspace{10mm} x=0,1,...
\end{equation}
with mean and variance:
\begin{equation}
\begin{aligned}
&E(X)= \frac{r(1-p)}{p} \hspace{8mm} \text{and}\hspace{8mm} 
V(X)= \frac{r(1-p)}{p^2}.
\end{aligned}
\label{eq:NBmeanvar}
\end{equation}
The normal approximation to the negative binomial is applied here for large $r$ and moderate $p$. This approximation is accurate under these conditions because both parameters, $r$ and $p$, affect the shape and symmetry of the negative binomial distribution. Hence, it closely resembles that of a normal distribution. This approximation enables us to meet the assumptions required of the Cauchy combination test for the individual tests. The Normal approximation to the negative binomial distribution becomes as follows:
\begin{equation}
\begin{aligned}
&X \approx N \left( E(X),V(X)\right).
\end{aligned}
\end{equation}




\section{Simulation study}\label{Simulation}
This simulation was designed to assess the type 1 error rates and the power of three different $p$-value combination methods: the Cauchy combination test (CCT), Fisher's test, and the MinP test using independent and correlated $p$-values obtained from the Normal approximation to the negative binomial distribution. In this context, the number of variables refers to the number of individual tests, or similarly $p$-values, denoted as $m$. We denote the sample size as $n$ and the number of simulations as $M$.
\subsection{Data generation}
Datasets for independent and correlated variables were simulated from the negative binomial (NB) distribution using the \texttt{R} software \citep{Rpackage}. The correlations were modelled using the Clayton and the Gumbel-Hougaard copulas to introduce the correlation among the variables. The dependence parameter, $\theta$, denotes the dependency strength between the variables. We considered $\theta$ values of 1, 3, and 5, which represent weak (independence in the Gumbel-Hougaard copula), moderate, and strong correlations in the lower tail of the Clayton copula or the upper tail of the Gumbel-Hougaard copula. First, we generated data using copulas with dimension $m$ and parameter $\theta$. Following this, the simulated unit variates from these copulas were transformed by quantile transformation to follow the negative binomial distribution.
\subsection{Type 1 error rate}
To evaluate the type 1 error rate, the data were simulated under the null hypothesis, $H_{0i}: \mu_i=\mu_{0i}, i=1,\ldots m$, using $M=10^5$ replications at $0.05$ and $0.01$ levels of significance. We generated datasets from negative binomial distribution with parameters $r$ and $p=0.5$, $NB(r, 0.5)$, each with a sample size $n=$30 under the null hypothesis $H_{0i}: \mu_i=r_{0i}, i=1,\ldots m$. We varied the following parameters: the number of variables $m$, the number of success parameter of the negative binomial distribution $r$, and copula parameter $\theta$. We calculated the Z test for each variable as $Z_i=\frac{\bar{X_i}-\mu_{0i}}{\sqrt{\frac{V(X_i)}{n}}}$, with a two-sided $p$-value given by $2[1-\Phi(|Z_i|)$, for $i=1,\ldots m$. Finally, we calculated the mean of the combined $p$-values that were below the prespecified significance level.

\begin{figure}[htb]
\centering
\includegraphics[width=0.49\textwidth]{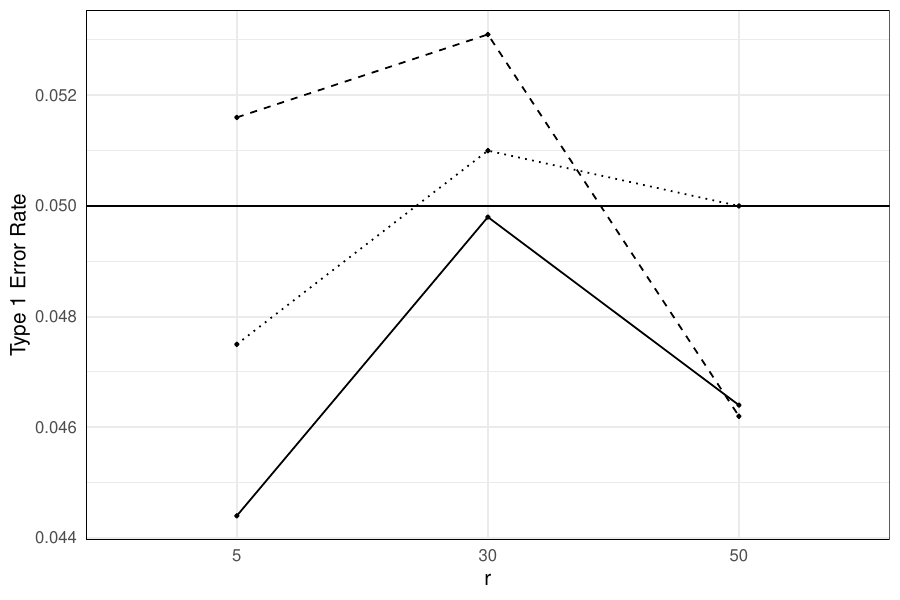}
\includegraphics[width=0.49\textwidth]{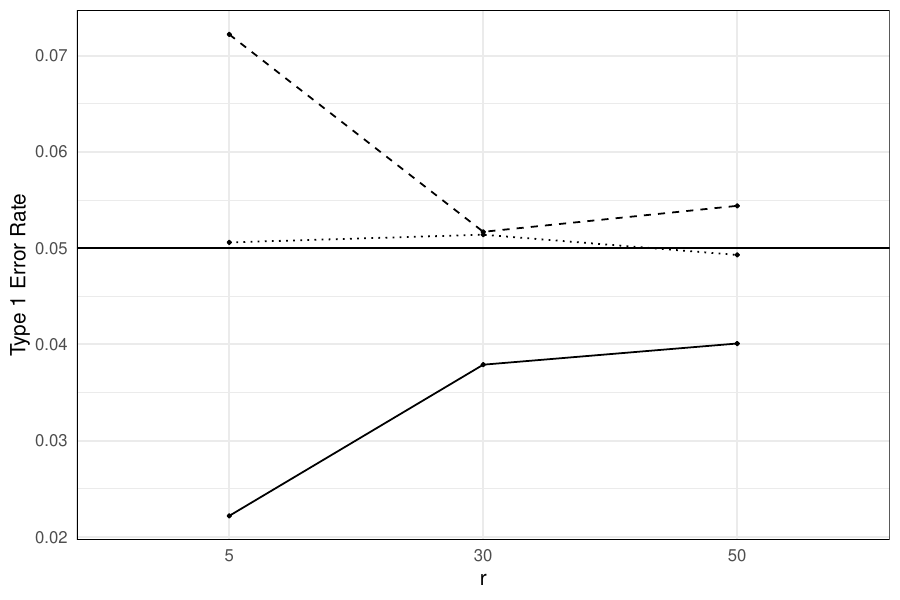}
\caption{Type 1 error rates for the Cauchy combination test (smooth line), Fisher's test (dotted line), and MinP test (dashed line) at the significance level $\alpha=0.05$ against the success parameter $r$, using $M=10,000$ replications. Datasets were simulated from $10$ (left) and $50$ (right) independent negative binomial variables with sample size $n=30$ and parameters $(r,0.5)$.}
\label{t1NBI}
\end{figure}

Figure~\ref{t1NBI} shows the results for the type 1 error rates against the parameter $r$ of the three combination methods in the independence case at the significance level $0.05$. Moreover, Table~\ref{tab:tableNBindependence} in Appendix~\ref{appendix:A} presents the type 1 error rates for different numbers of independent $p$-values, $r$, and significance levels. For correlated negative binomial data, the results are in Table~\ref{tab:tableNBClayton} and Table~\ref{tab:tableNBGumbel}.
\subsection{Power comparisons}
\begin{figure}[htb]
        \centering
\includegraphics[width=0.49\textwidth]{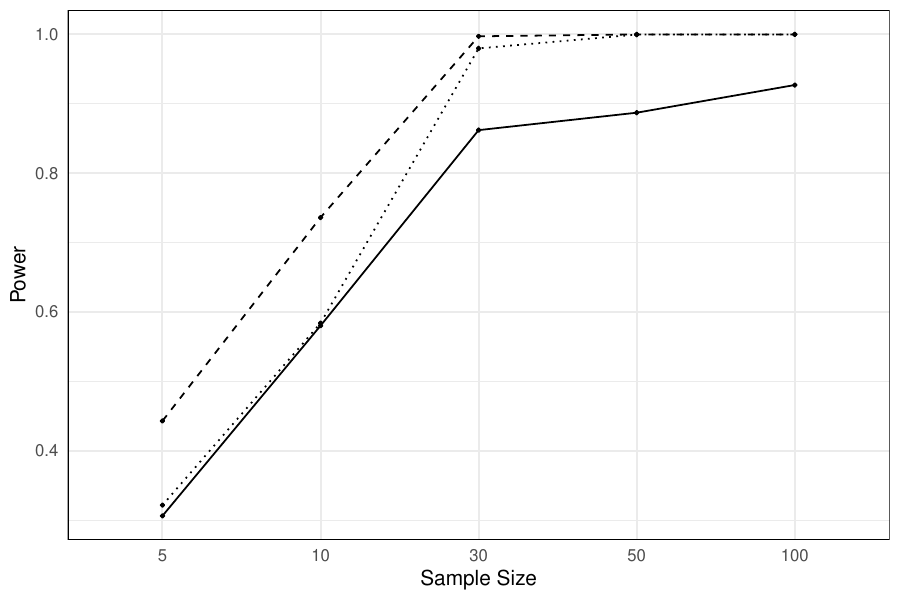}\\
\includegraphics[width=0.49\textwidth]{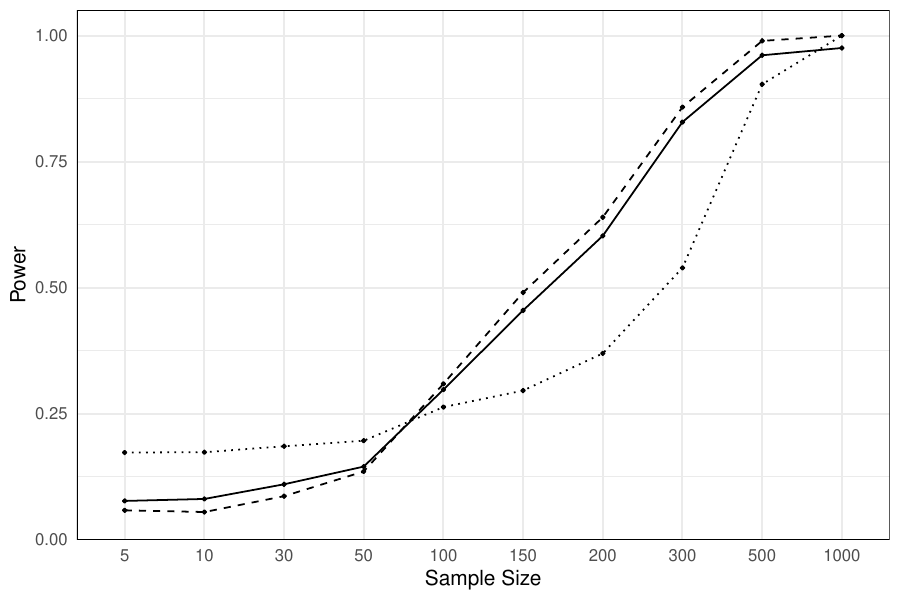}
\includegraphics[width=0.49\textwidth]{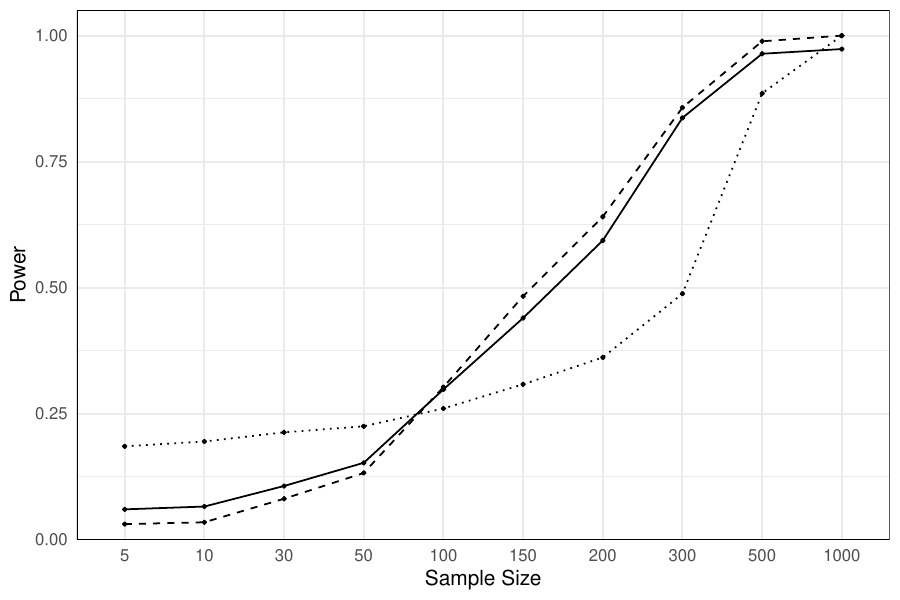}
    \caption{The power comparison of the Cauchy combination test (smooth line), Fisher's test (dotted line), and MinP test (dashed line) from negative binomial data in three different correlation structures: independence (top), the Clayton copula (bottom left), and the Gumbel-Hougaard copula (bottom right).}
    \label{fig:power2}
\end{figure}

The statistical powers of the three combination methods were compared in the presence of sparse signals. The evaluation was performed against the sample size and the correlation strength using different correlation structures.
Data were generated from a negative binomial distribution, consisting of nine variables with parameters $r=10$ and $p=0.5$, along with one variable with $r=11$ and $p=0.5$. We varied the correlation structures and evaluated the power against values of sample size $n=5,10,30,50,100,150,200,300,500,1000$ using $M=10,000$ replicated samples. The correlation structures included independence among all variables, and correlated variables modelled via the Clayton and the Gumbel-Hougaard copulas with $\theta$ equals $3$. Figure~\ref{fig:power2} illustrates the power of the three methods against the number of sample sizes.

\section{Results and discussion} \label{Results}
To evaluate the performance of the CCT, we conducted a simulation study to assess the type 1 error and power when the individual $p$-values were obtained from tests based on the Normal approximation to the negative binomial distribution. In addition, we studied the effect of the success parameter, the sample size, and the correlation structures.

Table~\ref{tab:tableNBindependence} in Appendix~\ref{appendix:A} presents the results for type 1 error rates when combining independent tests. To demonstrate the effect of the success parameter $r$, Figure~\ref{fig:power2} presents the results when the numbers of combined tests are $10$ and $50$. Across different values of the parameter $r$ and varying the number of tests $m$, the Fisher's test consistently controls the type 1 error well. When the number of tests is small, all methods manage to control false positives within different significance thresholds. However, the impact of the parameter $r$ on the type 1 error rates of the CCT and MinP tests is evident as $m$ increases. 

The MinP method exhibits relatively stable Type 1 error rates as $r$ increases while it remains conservative for the CCT. The CCT shows conservative type 1 error rates at the $0.05$ significance level, whereas at $0.01$, it is around $0.01$, except when $m=100$ and $r$ equals $5$ or $30$. As $r$ increases, the rate increases for the CCT especially for large $m$. For instance, when $m=50$ at $\alpha=0.05$, type 1 error rate increases from $0.0222$ to $0.0401$. 

Two types of correlation structures using the Clayton and the Gumbel-Hougaard copulas were introduced. The results are presented in Tables~\ref{tab:tableNBClayton} and~\ref{tab:tableNBGumbel}, Appendix~\ref{appendix:A}. The Fisher's test has the highest type 1 error rates, except in the Gumbel-Hougaard copula when $\theta$ is $1$, which represents the independence case. As expected, this is due to the violation of the independence assumption. Conversely, the MinP test tends to be more conservative in controlling type 1 errors across different copula structures and levels of dependence. Our findings show that the CCT outperforms the MinP test in controlling false positives, particularly when the dependence strength $\theta$ increases from $1$ to $5$ in the Gumbel-Hougaard copula.

For the CCT, the choice of copula model significantly affects controlling the type 1 error rates. In Table~\ref{tab:tableNBClayton}, as the parameters $r$ and $\theta$ grow, the CCT has slightly higher type 1 error rates for the different numbers of dependent tests through the Clayton copula. On the other hand, modelling tests based on the Gumbel-Hougaard copula show that as the correlation strength increases, the type 1 error rate decreases and becomes more controlled for highly correlated tests. For example, at $r=30$ and $\theta=5$, across different numbers of tests, the type I error rates for the CCT range from $0.0588$ to $0.062$, while in the Gumbel-Hougaard copula they range from $0.051$ to $0.0514$. 

Figure~\ref{fig:power2} compares the power of the three combination methods against the sample size. As expected, power is generally increasing as the sample size increases. The Fisher and MinP tests have higher power than the CCT when the combined individual tests are independent. The CCT and MinP exhibit comparable power in sparse signals and small effect sizes, regardless of the correlation structure. 

In this study, the success parameter $r$ in the negative binomial distribution has an implication in controlling type 1 error rates in hypothesis testing. As $r$ increases, the data become more symmetric and less overdispersed, resulting in a more accurate Normal approximation. The observed variations in the CCT's type 1 error rates for a large number of tests underline the importance of considering the effect of the parameter $r$ when combining multiple tests based on the Normal approximation. This underscores the need for caution to ensure the robustness and validity of statistical inferences. Furthermore, multivariate copulas with discrete marginal distributions do not have a unique copula representation \citep{Sklar1959}. As in the simulation study, modelling the count data with the Clayton copula may not uniquely capture the true relationships, potentially impacting the results of type 1 error rates. This notes the limitations and implications of non-uniqueness in copula modelling for count data \citep{Safari_Katesari_2020, Geenens+2020+417+440}.
\section{Conclusion}\label{Conclusion}
In this paper, we compared three $p$-value combination tests obtained from count data based on the Normal approximation to the negative binomial distribution. The Cauchy combination test (CCT) is powerful and robust against sparse alternatives under arbitrary dependence structures, but requires caution when combining independent or weakly dependent $p$-values. Factors such as the success parameter, the significance level, and the correlation model for count data could have a potential impact on the CCT.

\section*{Acknowledgements}
The first author acknowledges the financial support from King Abdulaziz University provided through the Saudi Cultural Bureau in London and the Ministry of Education in Saudi Arabia.
\section*{Conflict of interest}
The authors declare that they have no competing interests.



\newpage
\section{Appendices}

\appendix
\section{Type one error rate tables} \label{appendix:A}

\begin{table}[h!]
\tbl{Type 1 error rates for the CCT, Fisher, and MinP tests at different values of the level of significance $\alpha$ using $M=10,000$ replications. Datasets were simulated from $m$ independent negative binomial variables with sample size $n=30$ and parameters $(r,0.5)$.}
{\begin{tabular}{cccccc} \toprule
\multirow{2}{*}{\textbf{Number of tests (m)}} & \multirow{2}{*}{$\boldsymbol{r}$} & \multirow{2}{*}{\textbf{Test}} & \multicolumn{2}{c}{\textbf{Type one error rates at $\alpha$}} \\ 
\cmidrule{4-5}
 & & & $\boldsymbol{0.05}$ & $\boldsymbol{0.01}$ \\ \midrule
10 & 5 & CCT & 0.0444 & 0.0106\\
   &   & Fisher & 0.0475 & 0.0097 \\
   &   & MinP & 0.0516 & 0.0142 \\ \cmidrule{2-5}
   & 30 & CCT & 0.0498 & 0.0086 \\
   &    & Fisher & 0.0510 & 0.0103 \\
   &    & MinP & 0.0531 & 0.0096 \\\cmidrule{2-5}
   & 50 & CCT & 0.0464 & 0.0100  \\
   &    & Fisher & 0.0500 & 0.0120  \\
   &    & MinP & 0.0462 & 0.0095  \\ \cmidrule{1-5}
50 & 5 & CCT & 0.0222 & 0.0058  \\
   &   & Fisher & 0.0506 & 0.0110 \\
   &   & MinP & 0.0722 & 0.0188 \\ \cmidrule{2-5}
   & 30 & CCT & 0.0379 & 0.0066  \\
   &    & Fisher & 0.0514 & 0.0086  \\
   &    & MinP & 0.0517 & 0.0109 \\ \cmidrule{2-5}
   & 50 & CCT & 0.0401 & 0.0085  \\
   &    & Fisher & 0.0493 & 0.0096\\
   &    & MinP & 0.0544 & 0.0112  \\ \cmidrule{1-5}
100 & 5 & CCT & 0.0078 & 0.0019 \\
    &   & Fisher & 0.0495 & 0.0091 \\
    &   & MinP & 0.0665 & 0.0198 \\ \cmidrule{2-5}
   & 30 & CCT & 0.0230 & 0.0034 \\
    &    & Fisher & 0.0526 & 0.0113 \\
    &    & MinP & 0.0494 & 0.0097  \\ \cmidrule{2-5}
   & 50 & CCT & 0.0259 & 0.0061 \\
    &    & Fisher & 0.0469 & 0.0085  \\
    &    & MinP & 0.0484 & 0.0098 \\
\bottomrule
\end{tabular}}
\tabnote{CCT, Cauchy combination test; MinP, Minimum $P$-value test.}
\label{tab:tableNBindependence}
\end{table}

\begin{table}
\tbl{Type 1 error rates for the CCT, Fisher, and MinP tests at $0.05$ significance level using $M=10,000$ replications. Datasets were modeled using the Clayton copula $(\theta=1,3,5)$ and simulated from $m$ negative binomial variables with sample size $n=30$ and parameters $(r,0.5)$.}
{\begin{tabular}{cccccc} \toprule
\multirow{2}{*}{\thead{Number of tests \\ ($m$)}} & \multirow{2}{*}{$\boldsymbol{r}$} & \multirow{2}{*}{\thead{Test}} & \multicolumn{3}{c}{$\boldsymbol{\theta}$} \\ \cmidrule{4-6}
& & & 1 & 3 & 5 \\ \midrule
\multirow{6}{*}{10} & \multirow{3}{*}{5} & CCT & 0.0557 & 0.0604 & 0.059 \\
 &  & Fisher & 0.106 & 0.156 & 0.179 \\
 &  & MinP & 0.0489 & 0.0346 & 0.0294 \\ \cmidrule{2-6}
 & \multirow{3}{*}{30} & CCT & 0.0562 & 0.0603 & 0.0588 \\
 &  & Fisher & 0.1193 & 0.1679 & 0.1864 \\
 &  & MinP & 0.0418 & 0.0286 & 0.0251 \\ \midrule
\multirow{6}{*}{50} & \multirow{3}{*}{5} & CCT & 0.0476 & 0.067 & 0.066 \\
 &  & Fisher & 0.1841 & 0.2399 & 0.2574 \\
 &  & MinP & 0.0517 & 0.0317 & 0.0261 \\ \cmidrule{2-6}
 & \multirow{3}{*}{30} & CCT & 0.066 & 0.0665 & 0.06 \\
 &  & Fisher & 0.2574 & 0.2548 & 0.2648 \\
 &  & MinP & 0.0261 & 0.0196 & 0.0141 \\ \midrule
\multirow{6}{*}{100} & \multirow{3}{*}{5} & CCT & 0.0351 & 0.0713 & 0.0705 \\
 &  & Fisher & 0.221 & 0.2689 & 0.2834 \\
 &  & MinP & 0.0493 & 0.0311 & 0.0208 \\ \cmidrule{2-6}
 & \multirow{3}{*}{30} & CCT & 0.0594 & 0.0691 & 0.0625 \\
 &  & Fisher & 0.2353 & 0.281 & 0.2908 \\
 &  & MinP & 0.0376 & 0.0201 & 0.012 \\ \midrule
\multirow{6}{*}{500} & \multirow{3}{*}{5} & CCT & 0.0057 & 0.049 & 0.0654 \\
 &  & Fisher & 0.28 & 0.2948 & 0.3103 \\
 &  & MinP & 0.0611 & 0.0288 & 0.0194 \\ \cmidrule{2-6}
 & \multirow{3}{*}{30} & CCT & 0.03 & 0.065 & 0.0595 \\
 &  & Fisher & 0.29 & 0.3031 & 0.3192 \\
 &  & MinP & 0.0355 & 0.0155 & 0.0081 \\ \bottomrule
\end{tabular}}
\tabnote{CCT, Cauchy combination test; MinP, Minimum $P$-value test.}
\label{tab:tableNBClayton}
\end{table}

\begin{table}
\tbl{Type 1 error rates for the CCT, Fisher, and MinP tests at $0.05$ significance level using $M=10,000$ replications. Datasets were modeled using the Gumbel-Hougaard copula $(\theta=1,2,3)$ and simulated from $m$ negative binomial variables with sample size $n=30$ and parameters $(r,0.5)$.}
{\begin{tabular}{cccccc} \toprule
\multirow{2}{*}{\thead{Number of tests ($m$)}} & \multirow{2}{*}{$\boldsymbol{r}$} & \multirow{2}{*}{\thead{Test}} & \multicolumn{3}{c}{$\boldsymbol{\theta}$} \\ \cmidrule{4-6}
& & & 1 & 2 & 3 \\ \midrule
\multirow{6}{*}{10} & \multirow{3}{*}{5} & CCT & 0.0478 & 0.0529 & 0.0498 \\
 &  & Fisher & 0.0522 & 0.1932 & 0.2025 \\
 &  & MinP & 0.0534 & 0.0196 & 0.0138 \\ \cmidrule{2-6}
 & \multirow{3}{*}{30} & CCT & 0.0492 & 0.0539 & 0.051 \\
 &  & Fisher & 0.0533 & 0.1918 & 0.2014 \\
 &  & MinP & 0.0527 & 0.0193 & 0.0131 \\ \midrule
\multirow{6}{*}{50} & \multirow{3}{*}{5} & CCT & 0.0233 & 0.0536 & 0.0506 \\
 &  & Fisher & 0.0521 & 0.2707 & 0.2827 \\
 &  & MinP & 0.0669 & 0.0092 & 0.0053 \\ \cmidrule{2-6}
 & \multirow{3}{*}{30} & CCT & 0.0402 & 0.0568 & 0.0514 \\
 &  & Fisher & 0.0513 & 0.2722 & 0.2825 \\
 &  & MinP & 0.0528 & 0.0086 & 0.0044 \\ \midrule
\multirow{6}{*}{100} & \multirow{3}{*}{5} & CCT & 0.0104 & 0.0542 & 0.051 \\
 &  & Fisher & 0.0507 & 0.2948 & 0.3051 \\
 &  & MinP & 0.0633 & 0.0057 & 0.003 \\ \cmidrule{2-6}
 & \multirow{3}{*}{30} & CCT & 0.0236 & 0.0559 & 0.0511 \\
 &  & Fisher & 0.0521 & 0.2937 & 0.3046 \\
 &  & MinP & 0.0484 & 0.0061 & 0.0022 \\ \midrule
\multirow{6}{*}{500} & \multirow{3}{*}{5} & CCT & 0.0001 & 0.0554 & 0.0505 \\
 &  & Fisher & 0.0469 & 0.3255 & 0.3342 \\
 &  & MinP & 0.0868 & 0.0036 & 0.0014 \\ \cmidrule{2-6}
 & \multirow{3}{*}{30} & CCT & 0.0004 & 0.0555 & 0.0512 \\
 &  & Fisher & 0.0478 & 0.3259 & 0.3351 \\
 &  & MinP & 0.0557 & 0.0032 & 0.0011 \\ \bottomrule
\end{tabular}}
\tabnote{CCT, Cauchy combination test; MinP, Minimum $P$-value test.}
\label{tab:tableNBGumbel}
\end{table}

\end{document}